\documentclass{sf2a-conf2012}
\usepackage{graphicx}
\usepackage{hyperref}
\usepackage[]{natbib}  
\usepackage[cyr]{aeguill}
\usepackage{epstopdf}
\usepackage{textcomp}
\usepackage{amsmath}
\usepackage{txfonts}

\def\BibTeX{{\rm B\kern-.05em{\sc i\kern-.025em b}\kern-.08em
    T\kern-.1667em\lower.7ex\hbox{E}\kern-.125emX}}
\bibpunct{(}{)}{;}{a}{}{,}  

\newcommand{\mcal}{\mathcal}

\newcommand{\mrm}{\mathrm}

\renewcommand{\Im}{\mathcal I\!m \ }

\newcommand{\interval}[2]{ \textrm{\textlbrackdbl}#1,#2\textrm{\textrbrackdbl} }
 
\begin{document}

\TitreGlobal{SF2A 2012}


\title{The anelastic equilibrium tide in exoplanetary systems}

\runningtitle{The anelastic equilibrium tide in exoplanetary systems}

\author{F. Remus$^{1,2,3}$}
\author{S. Mathis$^{2}$}
\author{J.-P. Zahn$^{1}$}
\author{V. Lainey$^{}$}

\address{LUTH, Observatoire de Paris, CNRS, Universit\'e Paris Diderot, 92195 Meudon, France}
\address{Laboratoire AIM Paris-Saclay, CEA/DSM, CNRS, Universit\'e Paris Diderot, IRFU/SAp, 91191 Gif-sur-Yvette, France}
\address{IMCCE, Observatoire de Paris, CNRS, UPMC, USTL, 75014 Paris, France}

\setcounter{page}{237}

\index{Remus, F.}
\index{Mathis, S.}
\index{Zahn, J.-P.}
\index{Lainey, V.}


\maketitle


\begin{abstract}

Earth-like planets have anelastic mantles, whereas giant planets may have anelastic cores. 
As for the fluid parts of a body, the tidal dissipation of such solid regions, gravitationally perturbed by a companion body, highly depends on its internal friction, and thus on its internal structure. 
Therefore, modelling this kind of interaction presents a high interest to provide constraints on planet interiors, whose properties are still quite uncertain.

Here, we examine the equilibrium tide in the solid central region of a planet, taking into account the presence of a fluid envelope. 
We first present the equations governing the problem, and show how to obtain the different Love numbers that describe its deformation.
We discuss how the quality factor Q depends on the rheological parameters, and the size of the core.

Taking plausible values for the anelastic parameters, and examinig the frequency-dependence of the solid dissipation, we show how this mechanism may compete with the dissipation in fluid layers, when applied to Jupiter- and Saturn-like planets. 
We also discuss the case of the icy giants Uranus and Neptune.
\end{abstract}



\section{Introduction}

Once a planetary system is formed, its dynamical evolution is governed by gravitational interactions between its components, be it a star-planet or planet-satellite interaction. 
By converting kinetic energy into heat, the tides pertub their orbital and rotational properties, and the rate at which the system evolves depends on the physical properties of tidal dissipation. 
Therefore, to understand the past history and predict the fate of a binary system, one has to identify the dissipative processes that achieve this conversion of energy. 
Planetary systems display a large diversity of planets, with telluric planets having anelastic mantles and giant planets with possible anelastic cores \citep{2007ARA&A..45..397U}. 
Since the tidal dissipation is closely related with the internal structure, one has to investigate its effects on each kind of materials that may compose a planet. 
Studies have been carried out on tidal effects in fluid bodies such as stars and envelopes of giant planets \citep{2004ApJ...610..477O, 2007ApJ...661.1180O, 2009MNRAS.396..794O, 2012A&A...541A.165R}. 
However, the planetary solid regions, such as the mantles of Earth-like planets or the rocky cores of giant planets may also contribute to tidal dissipation \citep[see for example][]{2012ApJ...746..150E, 2012A&A...541A.165R}.
We explore here the tidal dissipation in these solid parts of planets.

\section{The system}

\paragraph*{Two-layer model. --}

\begin{figure}[!htb]
 \centering
 \begin{minipage}[m]{0.52\linewidth}
	\includegraphics[width=0.8\linewidth]{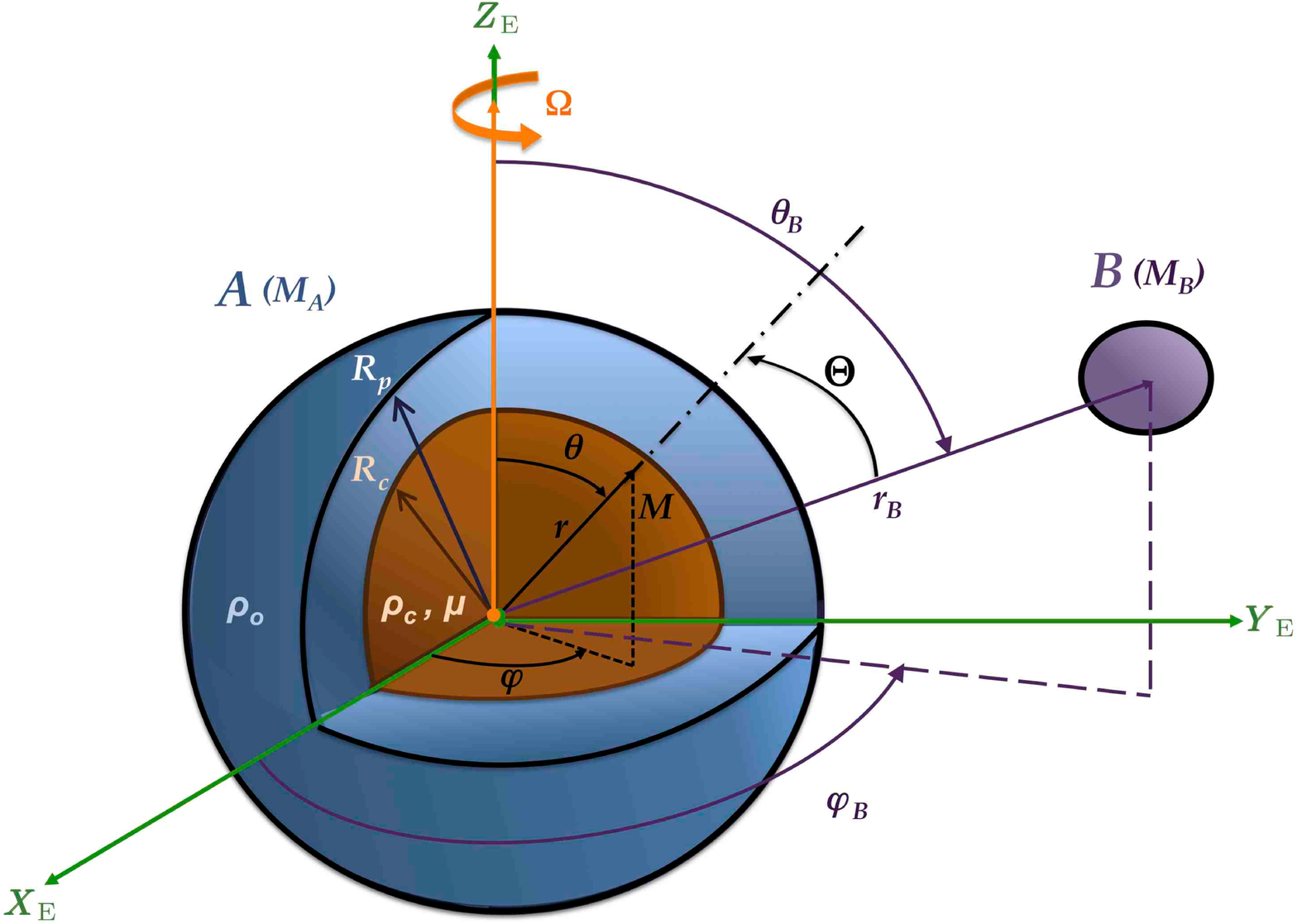}
 \end{minipage}
 \begin{minipage}[m]{0.38\linewidth}
 \centering
 		\includegraphics[width=\linewidth]{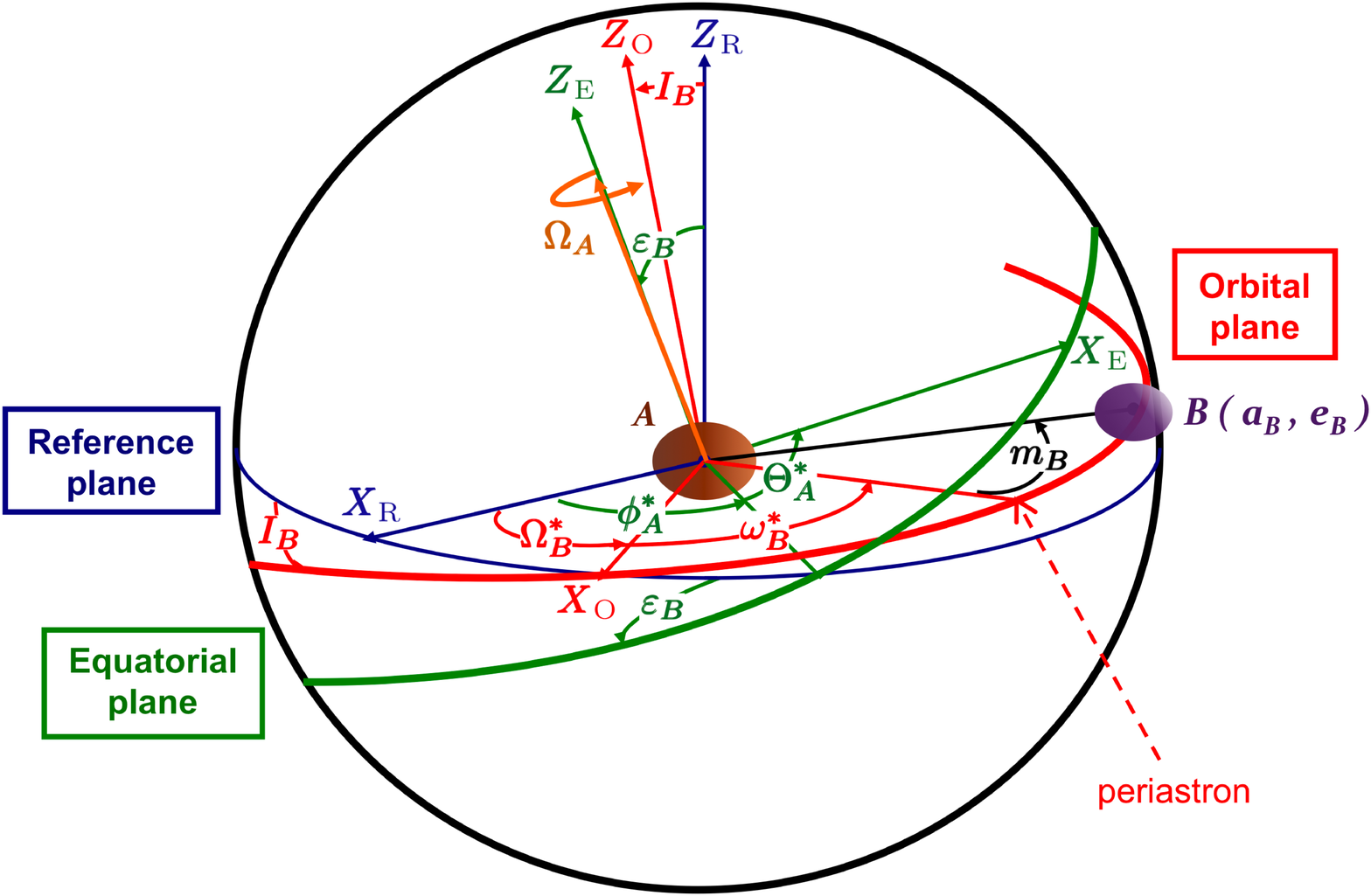}
 \end{minipage}
 \caption{
 {\bf Left:} the system is composed by a two-layer main component A, with an homogeneous and incompressible solid core and an homogeneous static fluid envelope, and a point-mass perturber B orbiting around A.
 {\bf Right:} B is supposed to move on an elliptical orbit, inclined with respect to the inertial reference plane $\left(X_R,Y_R \right)$. The equatorial plane of A $\left(X_E,Y_E \right)$ is also inclined with respect to this same reference plane.
 }
 \label{remus_fig1}
\end{figure}

We will consider as a model a two-bodies system where the component A, rotating at the angular velocity $\Omega$, has a viscoelastic core of  shear modulus $\mu$, made of ice or rock, surrounded by a fluid envelope, such as an ocean, streching out from core's surface (of mean radius $R_c$) up to planet's surface (of mean radius $R_p$).
Both core and envelope are considered homogeneous, with constant density $\rho_c$ and $\rho_o$ respectively.
This model is represented on the left panel of Fig. \ref{remus_fig1}.

\paragraph*{Configuration. --}

We undertake to describe the tide exerted by B (of mass $m_B$) on the solid core of A, when moving in an elliptic orbit around A, with eccentricity $e$, at the mean motion $\omega$.
Since no assumption is made on the B's orbit, we need to define an inclination angle $I$ to determine the position of the orbital spin of B with respect to the total angular momentum of the system  (in the direction of $Z_R$) wich defines an inertial reference plane $\left(X_R,Y_R \right)$, perpendicular to it.
The spin axis of A then presents an obliquity $\varepsilon$ with respect to $Z_R$.
Refer to the right panel of Fig. \ref{remus_fig1} for a synthetic representation of the system configuration.

To treat the complexity of the two-layer problem, we follow the methodology of \citet{1979Icar...37..310D}.

\section{Tidal dissipation of the core in the case of a two-layer planet}

\paragraph*{Definition. --}

The tidal perturbation exerted by B on the solid core of A results on one hand in its deformation, and on the other hand in the dissipation of the tidal energy into heat leading to a lag angle $\delta$ between the line of centers and the tidal bulge.
This process can be modeled by the complex second-order Love number $\tilde{k}_2$ defined as the ratio of the perturbed gravific response potential $\Phi'$ over the tidal potential $U$ (\citealp{1954JAP....25.1385B}, \citealp[see also][]{TheseTobie2003} and \citealp{2009ApJ...707.1000H}). 
Its real part represents then the purely elastic deformation of the potential of the core ($\Phi'$) while its imaginary part accounts for its anelastic tidal dissipation.

In practical calculations, we first have to develop $U$ (and therefore $\Phi'$) on spherical harmonics ($Y_2^m$), each term having a wide range of tidal frequencies $\sigma_{2,m,p,q} = (2-2p+q) \, \omega - m \Omega$, for $(m,p,q) \in \interval{-2}{2} \times \interval{0}{2} \times \mathbb{Z}$, resulting from the expansion of $U$ on the Keplerian elements using the Kaula transform (\citealp{1962AJ.....67..300K}, \citealp[see also][]{2009A&A...497..889M}).
Thus, the complex Love number $\tilde{k}_2$ depends on the tidal frequency and the rheology of the core, and so does the {\it quality factor} $Q$ which quantifies the tidal dissipation \citep[see for example][]{TheseTobie2003} 
\begin{equation}
	Q^{-1}(\bar{\mu} , \sigma_{2,m,p,q}) = - \frac{\Im{\tilde{k}_2(\bar{\mu} , \sigma_{2,m,p,q})}}{\left| \tilde{k}_2(\bar{\mu} , \sigma_{2,m,p,q}) \right|}   \:,
	\:\: \text{where } \: \tilde{k}_2(\bar{\mu} , \sigma_{2,m,p,q}) = \frac{\Phi'(\bar{\mu} , \sigma_{2,m,p,q})}{U(\sigma_{2,m,p,q})}  = \left| \tilde{k}_2 \right| \, e^{ -i \left[ 2\delta (\bar{\mu} , \sigma_{2,m,p,q}) \right] } \,,
	\label{Q_def}
\end{equation}
where the quantity $\bar{\mu} \equiv \bar{\mu}_1 + \pmb i \, \bar{\mu}_2 = \frac{19 \, \mu}{2 \, \rho_c \, g_c \, R_c}$ is the complex effective shear modulus, linked with the anelasticity (and thus the rheology) of the planet's core and its gravity $g_c$.

\paragraph*{Case of a two-layer planet. --}

Acting as an overload on the solid core, the fluid shell, deformed by the tide, modifies both the tidal deformation and dissipation of the core.
The second order Love number $\tilde{k}_2$ takes then a different form than in the fully-solid case
\begin{equation}
\label{k2i'}
	\tilde{k}_2(\bar{\mu} , \sigma_{2,m,p,q}) = \frac{1}{ \left(B+\bar{\mu}_1 \right)^2 + \bar{\mu}_2^2 }
		\times \left\lbrace  \left[  \left(B+\bar{\mu}_1 \right) \, \left( C+\frac{3}{2\alpha} \, \bar{\mu}_1 \right) + \frac{3}{2\alpha} \, \bar{\mu}_2^2 \right] 
	     - \pmb i {A \, D \, \bar{\mu}_2}  \right\rbrace \:,
\end{equation}
where $\alpha$, $A$, $B$, $C$ and $D$ account for the planet's internal structure through the ratios of radii $\frac{R_c}{R_p}$ and densities $\frac{\rho_o}{\rho_c}$.\\

Thus, the expression of the associated tidal dissipation rate
\begin{equation}
	Q(\bar{\mu} , \sigma_{2,m,p,q}) = \sqrt{ 
		1 
		+ \frac{9 \, \bar{\mu}_2(\sigma_{2,m,p,q})^2}{4 \alpha^2 \, A^2 \, D^2} 
			\left[
				1 + 
				\frac{\left( B + \bar{\mu}_1(\sigma_{2,m,p,q}) \right) \, \left( \frac{2\alpha C}{3} + \bar{\mu}_1(\sigma_{2,m,p,q}) \right)}
				     { \bar{\mu}_2(\sigma_{2,m,p,q})^2 }
			\right]^2
		}
\end{equation}
depends on the core's parameters (its size, density and rheological parameters) and the tidal frequency.
Moreover, to derive this expression of $Q$, no assumption has been made on the rheology of the core, except that it is linear under the small tidal perturbations (i.e. core's material obeys the Hooke's law).
Hence, it is valid for any linear rheological model.

\paragraph*{Comparison with observations. --}
To confront our model with observations, we need to introduce the global dissipation factor, corresponding to a \emph{rescaling} of the previous one to the planet surface and thus involving the second-order Love number at the surface of the planet
\begin{equation}
	Q_\mrm{eff} = \left( \frac{R_p}{R_c} \right)^5 \times \left| \frac{\tilde{k}_2(R_p)}{\tilde{k}_2(R_c)} \right| \times Q  \:.
\end{equation}

Moreover, we need to choose a model to represent the way the core's material responds to the tidal perturbation, i.e. a \emph{rheological model}.
Thus, from now on, we assume that the core behaves like a Maxwell body \citep[see, for example,][]{TheseTobie2003}.

\section{Application to giant planets}

\paragraph*{Application to gas giants. --}

\begin{figure}[!htb]
 \centering
 \begin{minipage}[m]{0.445\linewidth}
	\includegraphics[width=0.75\linewidth]{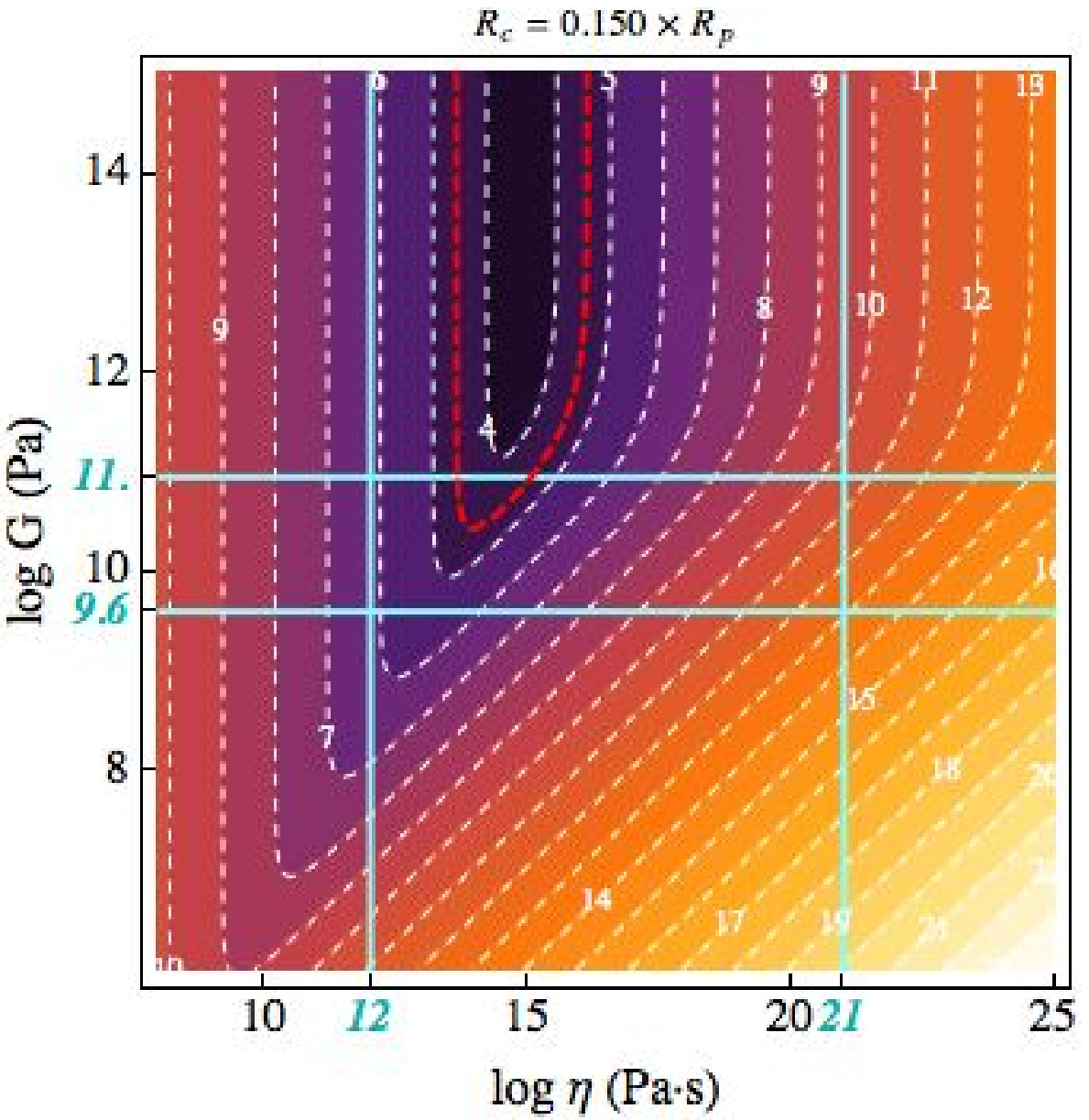}
 \end{minipage}
 \begin{minipage}[m]{0.445\linewidth}
	\includegraphics[width=0.7\linewidth]{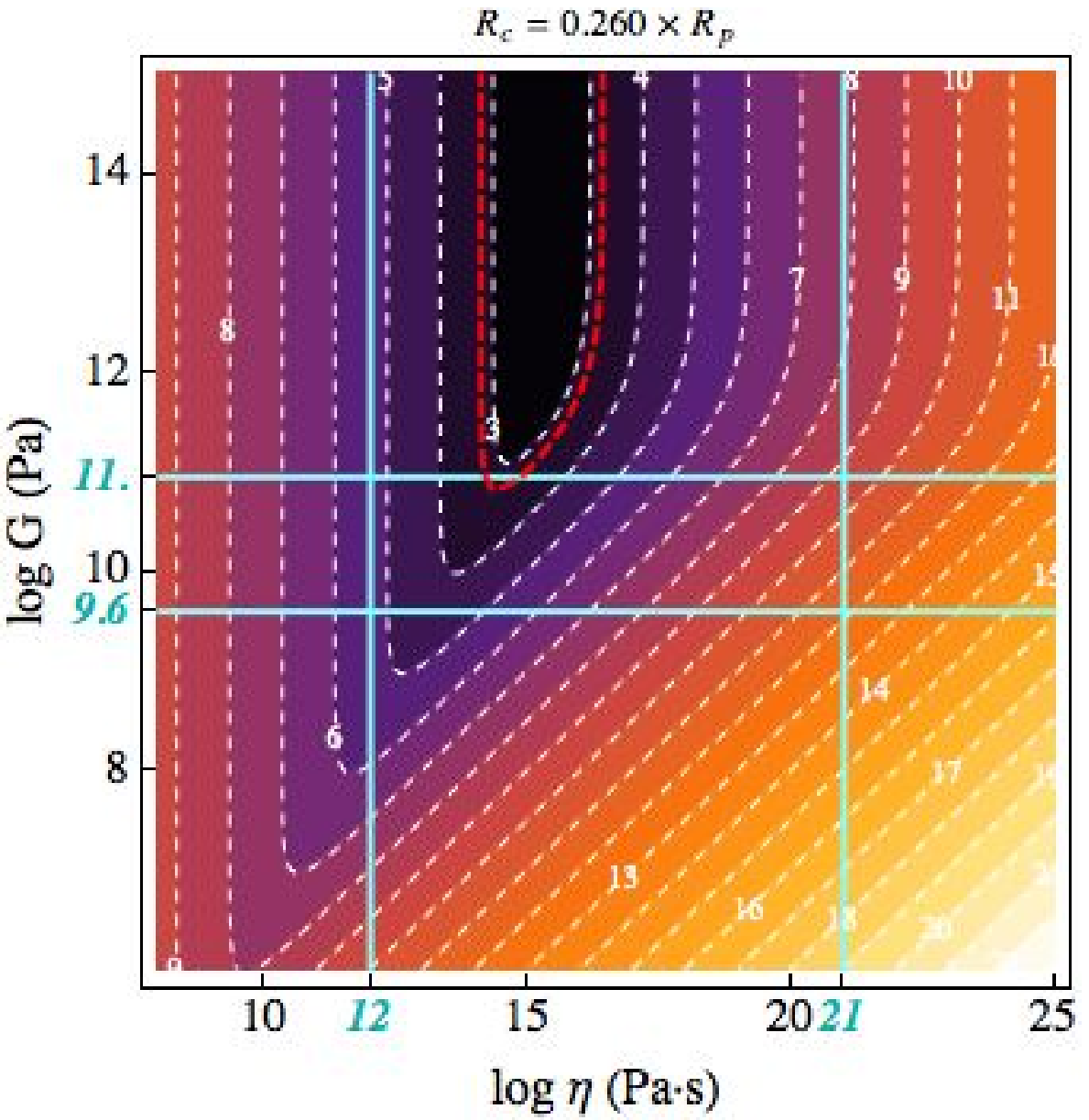}
 \end{minipage}
 \begin{minipage}[m]{0.1\linewidth}
	\includegraphics[width=0.7\linewidth]{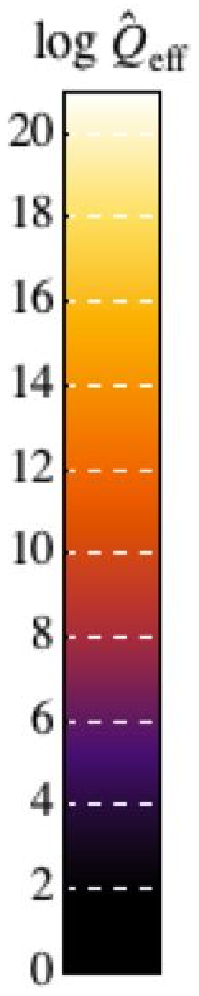}
 \end{minipage}
	\caption{
	Dissipation quality factor $Q_\mrm{eff}$ as a function of the viscoelastic parameters $G$ and $\eta$, of a two-layer gas giant, using the Maxwell model.
 {\bf Left:} for a Jupiter-like planet at the tidal frequency of Io.
 {\bf Right:} for a Saturn-like planet at the tidal frequency of Enceladus.
 The red dashed line indicates the value of $\hat{Q}_\mrm{eff}=\{(3.56 \pm 0.56) \times 10^4,(1.682 \pm 0.540) \times 10^3\}$ (for Jupiter and Saturn, respectively) determined by \citet{2009Natur.459..957L,2012ApJ...752...14L}.
  The blue lines corresponds to the lower and upper limits of the reference values taken by the viscoelastic parameters $G$ and $\eta$ for an unknown mixture of ice and silicates.
  We assume the values of ${R_p = \{10.97 , 9.14\}}$ (in units of $R_{\mcal\Phi}$), ${M_p = \{317.8, 95.16\}}$  (in units of $M_{\mcal\Phi}$), ${R_c = \{0.15, 0.26\} \times R_p}$, and ${M_c = \{6.41, 18.65\}}$  (in units of $M_{\mcal\Phi}$).
 \label{remus_fig2}
 }
\end{figure}

Using astrometric data covering more than a century, \citet{2009Natur.459..957L,2012ApJ...752...14L} succeeded in determining from observations the tidal dissipation in Jupiter and Saturn: namely, ${Q_\mrm{Jupiter} = (3.56 \pm 0.56) \times 10^4}$ \citep{2009Natur.459..957L}, and ${Q_\mrm{Saturn} = (1.682 \pm 0.540) \times 10^3}$ determined by \citet{2012ApJ...752...14L}. Note that such high dissipation is required by the formation scenario of Saturn's system of \citet{2011Icar..216..535C}, in which the mid-sized satellites are formed at the edge of the rings.
These values, which seem to be in agreement with other observations related to Jupiter's and Saturn's systems (see the corresponding references cited just above), are lower of up to one order of magnitude than what was expected by previous formation scenarios \citep[see, for example,][]{1981Icar...47....1Y,1983ASSL..106...19S}, and even lower than what the most up-to-date models of fluid tidal dissipation predict \citep[see, for example,][]{2004ApJ...610..477O,2005ApJ...635..688W}.
Then, the question arises on the role of the possible solid central regions as sources of dissipation.
Since the composition of giant planets cores is poorly constrained \citep{2005AREPS..33..493G}, we explore in Fig. \ref{remus_fig2} the tidal dissipation of Jupiter's and Saturn's core for a large range of values of the viscoelastic parameters considering the Maxwell rheological model.
The other parameters (planet and core sizes and masses) are indicated in the legend. 

In \citeyear{2004ApJ...610..477O}, \citeauthor{2004ApJ...610..477O} studied tidal dissipation in rotating giant planets resulting from the excitation by the tidal potential of inertial waves in the convective region.
Taking into account the presence of a solid core as a boundary condition for the reflexion of inertial waves, they obtained 
a quality factor $Q_\mrm{eff} \approx 5 \times 10^5$.

The present two-layer model proposes an alternative process that may reach the values observed in \citet{2009Natur.459..957L,2012ApJ...752...14L}, depending on the viscosity $\eta$ and the stiffness $G$.

To explain the tidal dissipation observed in the gas giant planets of our Solar System, all processes have to be taken into account.

\paragraph*{Application to ice giants. --}

As in gas giants, the standard three-layer models for the interior structure of ice giants predict the presence of a solid rocky core \citep[see, for example,][]{1991Sci...253..648H, 1995P&SS...43.1517P, 1999Sci...286...72G}.
But it still remains an incertitude on the phase state of the intermediate "icy" layer located between the rocky core and the convective atmosphere.
Considering recent three-dimensional simulations of Neptune's and Uranus' dynamo that predict that this region is a stably stratified conductive fluid one \citep{2004Natur.428..151S, 2006Icar..184..556S}, \citet{2011Icar..211..798R} studied the electric conductivity of warm dense water taking into account the phase diagram of water.
Their results infer that part of this shell is in the superionic state, i.e. a two-component system of both a conducting proton fluid and a crystalline oxygen solid, and extends to about 0.42-0.56 of the planet radius.
Thus, it seems reasonable to assume for our two-layer model that the solid central region extends from the rocky core surface up to somewhere in the superionic shell.

\begin{figure}[!htb]
 \centering
 \begin{minipage}[m]{0.87\linewidth}
 \begin{minipage}[r]{\linewidth}
	\includegraphics[width=0.95\linewidth]{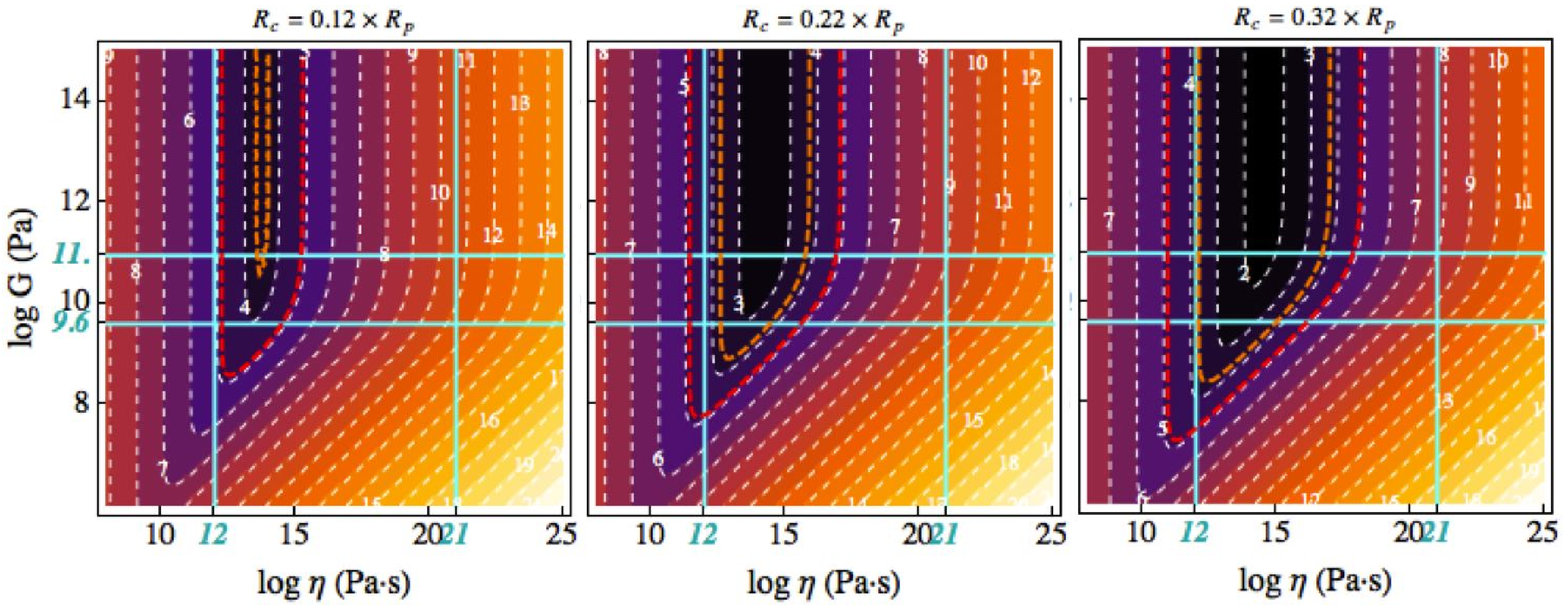}
 \end{minipage}
 \begin{minipage}[r]{\linewidth}
	\includegraphics[width=0.95\linewidth]{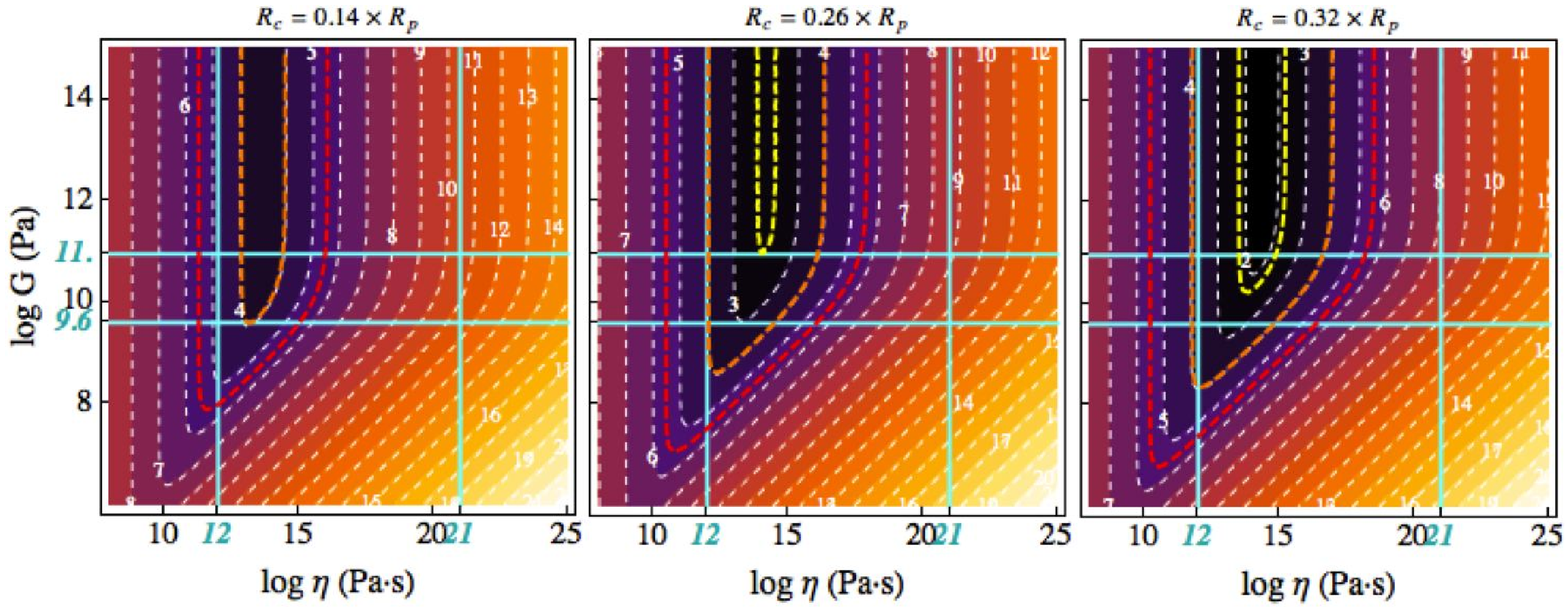}
 \end{minipage}
 \end{minipage}
 \begin{minipage}[m]{0.07\linewidth}
	\includegraphics[width=\linewidth]{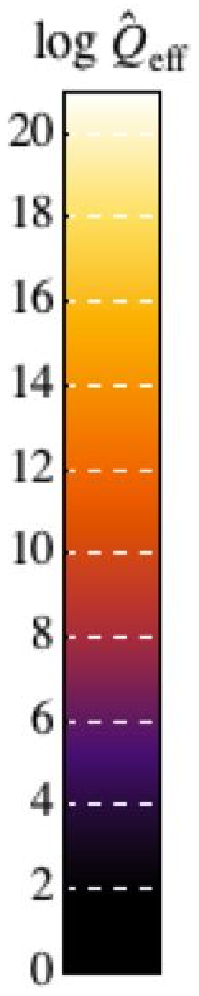}
 \end{minipage}
	\caption{
	Dissipation quality factor $Q_\mrm{eff}$ as a function of the viscoelastic parameters $G$ and $\eta$, of a two-layer ice giant, using the Maxwell model.
 {\bf Top:} for a Uranus-like planet at the tidal frequency of Miranda, with three different core sizes ${R_c = \{0.12, 0.22, 0.32\} \times R_p}$.
 {\bf Bottom:} for a Neptune-like planet at the tidal frequency of Triton, with three different core sizes ${R_c = \{0.14, 0.26, 0.32\} \times R_p}$.
 The red and orange dashed lines indicate, respectively, the lowest and highest values of $\hat{Q}_\mrm{eff}$ from formation scenarios: ${\hat{Q}_\mrm{eff} = \{5 \times 10^3, 7.2 \times 10^4 \}}$ for Uranus \citep{1977Icar...32..443G, 1966Icar....5..375G} and ${\hat{Q}_\mrm{eff} = \{9 \times 10^3, 3.3 \times 10^5 \}}$ for Neptune \citep{2008Icar..193..267Z, 1992Icar...99..390B}.
 The yellow dashed line indicates the value of $\hat{Q}_\mrm{eff}$ from a study of Neptune's internal heat: ${\hat{Q}_\mrm{eff}=1.7 \times 10^2}$ \citep{1974ApJ...193..477T}.
  The blue lines corresponds to the lower and upper limits of the reference values taken by the viscoelastic parameters $G$ and $\eta$ for an unknown mixture of ice and silicates.
  We assume the values of ${R_p = \{3.98 , 3.87\}}$ (in units of $R_{\mcal\Phi}$) and ${M_p = \{14.24, 16.73\}}$  (in units of $M_{\mcal\Phi}$).
  The core mass is obtained by integration of the density profiles of \citet{2011ApJ...726...15H} up to a given core size.
 \label{remus_fig3}
 }
\end{figure}

We explore in Fig. \ref{remus_fig3} the tidal dissipation of Uranus' and Neptune's core for a large range of values of the viscoelastic parameters, considering the Maxwell rheological model, for different core sizes.

\section{Dynamical evolution}

Due to dissipation, the tidal torque has non-zero average over the orbit, and it induces an exchange of angular momentum between each component and the orbital motion. 
This exchange governs the evolution of the semi-major axis $a$, the eccentricity $e$ of the orbit, the inclination $I$ of the orbital plane, of the obliquity $\varepsilon$ and that of the angular velocity of each component \citep[see for example][]{2009A&A...497..889M}. 
Depending on the initial conditions and on the planet/star mass ratio, the system evolves either to a stable state of minimum energy (where all spins are aligned, the orbits are circular and the rotation of each body is synchronized with the orbital motion) or the planet tends to spiral into the parent star.

\section{Conclusion}

Our evaluations reveal a much higher dissipation in the solid cores of planets than that found by \cite{2004ApJ...610..477O} for the fluid envelope of a planet having a small solid core. 
These results seem to be in good agreement with observed properties of Jupiter's and Saturn's system \citep{2009Natur.459..957L,2012ApJ...752...14L}.
In the case of the ice giants Uranus and Neptune, too much uncertainties remain on internal structure to give an order of magnitude, other than a minimum value, of tidal dissipation in the solid regions, which constitutes a first step in the tudy of such planets.


\begin{acknowledgements}
{
This work was supported in part by the Programme National de Plan\'etologie (CNRS/INSU), the EMERGENCE-UPMC project EME0911, and the CNRS {\it Physique th\'eorique et ses interfaces} program.
}
\end{acknowledgements}



\bibliographystyle{aa}  
\bibliography{RMZL12_sf2a.bib} 

\end{document}